\documentclass[prl, superscriptaddress, showpacs,floatfix, letterpaper, twocolumn]{revtex4} 
\usepackage{hyperref}
\usepackage{amssymb}
\usepackage{amsmath}
\usepackage{float}
\usepackage{graphicx}
\usepackage{epsfig}
\usepackage{epstopdf}
\usepackage[usenames]{color}

\begin{document}

\author{Sebastian Fuchs} 
\affiliation{Institut f\"{u}r theoretische Physik, Georg-August-Universit\"{a}t G\"{o}ttingen, 37077 G\"{o}ttingen, Germany}

\author{Emanuel Gull} \affiliation{Department of Physics, Columbia University, New York, NY 10027, USA} 

\author{Lode Pollet} 
\affiliation{Physics Department, Harvard University, Cambridge, Massachusetts 02138, USA}
 \affiliation{Theoretische Physik, ETH Zurich, 8093 Zurich, Switzerland}

\author{Evgeny Burovski} \affiliation{LPTMS, CNRS and Universit\'{e} Paris-Sud, UMR8626, B\^{a}timent 100, 91405 Orsay, France}
\author{Evgeny Kozik} \affiliation{Theoretische Physik, ETH Zurich, 8093 Zurich, Switzerland}

\author{Thomas Pruschke} 
\affiliation{Institut f\"{u}r theoretische Physik, Georg-August-Universit\"{a}t G\"{o}ttingen, 37077 G\"{o}ttingen, Germany}

\author{Matthias Troyer} \affiliation{Theoretische Physik, ETH Zurich, 8093 Zurich, Switzerland}

\title{Thermodynamics of the 3D Hubbard model on approach to the N\'{e}el transition}

\date{\today}

\hyphenation{}

\begin{abstract}
  We study the thermodynamic properties of the 3D Hubbard model for temperatures down to the N{\'e}el
  temperature using cluster dynamical mean-field theory. In particular we calculate the energy, entropy,  density, double occupancy and nearest-neighbor spin correlations as a function of chemical potential, temperature and repulsion strength. To make
  contact with cold-gas experiments, we also compute properties of the system subject to an
  external trap in the local density approximation. We find that an
  entropy per particle $S/N \approx 0.65(6)$ at $U/t=8$ is sufficient to achieve a N\'eel state in the center of the trap, substantially higher than the entropy  required in a homogeneous system. Precursors to antiferromagnetism can clearly be observed in
  nearest-neighbor spin correlators.
\end{abstract}


\maketitle

The Hubbard model~\cite{Hubbard63} remains one of the cornerstone
models in condensed matter physics, capturing the essence of strongly
correlated electron physics relevant to high-temperature
superconductors \cite{PWA} and correlation driven insulators
\cite{Imada98}. While qualitative features of the phase diagram are known from
analytical approximations, controlled quantitative studies in the
low-temperature regimes relevant for applications are not readily
tractable with tools presently available. A recent program that aims
to implement the Hubbard model in a cold gases
experiment~\cite{Koehl05} has led to experimental signs of the Mott
insulator~\cite{Joerdens08, Schneider08}. Modeling by dynamical mean
field theory (DMFT)~\cite{Schneider08, DeLeo08} and high-temperature series
expansions~\cite{Scarola09} resulted in temperature and entropy
estimates and an error budget~\cite{Joerdens10}. 
A major experimental
achievement will be the detection of the antiferromagnetic phase, for
which the slow and ill-understood equilibration rates, the limited
number of detection methods, and inherent cooling problems will
have to be overcome.  

Experimental progress has also sparked
interest in simulations of the 3D Hubbard model where new algorithms have been developed to treat the Hubbard model, such as the real-space
DMFT~\cite{PotthoffNolting,Gorelik}
%
%
%
%
or diagrammatic Monte Carlo~\cite{Kozik10} studies. Similar to the case of bosons, where synergy between experiment and simulation has led to quantitative understanding of experiments \cite{Trotzky09}, accurate results for the thermodynamics of the 3D Hubbard model will also be crucial for validation, calibration and thermometry of fermionic experiments. A crucial role
is played by the entropy, since these experiments form isolated systems where ideally the parameters are changed adiabatically (and not isothermally).  

In this Letter we provide the full thermodynamical equation of state of the Hubbard model -- in particular the entropy, energy, density, double occupancy and spin correlations -- 
for interactions $U$ up to the bandwidth $12t$ on approach to the
N{\'e}el temperature $T_N$ by performing controlled large-scale cluster
dynamical mean field calculations and extrapolations to the infinite system size limit, as well as determinantal diagrammatic Monte Carlo (DMC) simulations at half filling.
We then use this information to calculate the entropy per particle required for experiments on ultracold atomic gases in optical lattices to reach a N{\'e}el state in the trap center.
We finally show that the nearest-neighbor spin correlation
function contains clear precursors for antiferromagnetism that may already be
detectable in current generation experiments and that are useful
for thermometry (more so than measurements of the double occupancy) close to $T_N$.

The Hubbard model is defined by its Hamiltonian
\begin{equation}
  \hat{H} = -t \sum_{\langle i,j\rangle, \sigma}    \hat{c}_{i \sigma}^{\dagger} \hat{c}_{j \sigma}^{\phantom{\dagger}}
  + U \sum_i  \hat{n}_{i \uparrow} \hat{n}_{i \downarrow} - \sum_{i, \sigma}  \mu_i \hat n_{i \sigma}\; ,
  \label{FH}
\end{equation}
where $\hat{c}_{i \sigma}^{\dagger}$ creates a fermion with spin
component $\sigma=\uparrow, \downarrow$ on site $i$, $ \hat{n}_{i
  \sigma} = \hat{c}_{i \sigma}^{\dagger}\hat{c}_{i
  \sigma}^{\phantom{\dagger}}$, $\langle \dots \rangle$ denotes
summation over neighboring lattice sites, $t$ is the hopping amplitude,
$U$ the on-site repulsion, and $\mu_i = \mu - V(\vec{r}_i)$ with $\mu$ the
chemical potential and $V(\vec{r}_i)$ the confining potential at the location of the $i$-th lattice site.  We will set $V(\vec{r})=0$ in all calculations and consider realistic traps later on.

Our numerical approach is a cluster generalization of dynamical mean field theory \cite{Maier05}. In cluster DMFT the self
energy is approximated by $N_c$ momentum-dependent basis functions
$\phi_K(k)$: $\Sigma(k, \omega) \approx \sum_K^N \phi_K(k)
\Sigma_K(\omega)$. The exact problem is recovered for $N_c \rightarrow
\infty$. Within the dynamical cluster approximation (DCA)\cite{Hettler98}
used here, $\phi_K(k)$ are piecewise constant over momentum patches.
The DMFT method~\cite{Georges96} is the $N_c=1$ cluster
approximation where $\Sigma=\Sigma(\omega)$ and no momentum dependence
is retained.

\begin{figure}[t]
  \includegraphics[angle=0, width=0.9\columnwidth]{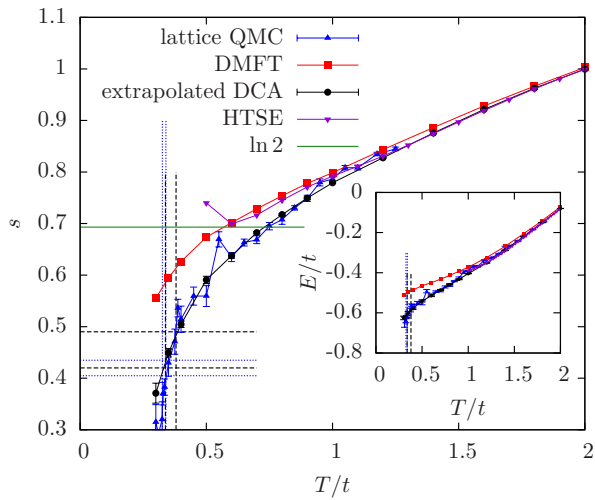}
  \caption{(color online) Entropy per lattice site $s$ of the Hubbard
    model as a function of temperature $T/t$, for $U/t=8$, at half
    filling.  Shown with dashed vertical lines is $T_N$ from Ref.  \cite{Kent05} (black) and with dotted lines according to DMC  (blue). Shown with dashed horizontal lines is the entropy per
    lattice site $s$ at $T_N$~\cite{Kent05} (black), with dotted lines according to DMC (blue); also $\log(2)$ is shown as a full horizontal
    line (green). The inset shows the energy $E$ per lattice site. }
  \label{fig:entropy_U8_0}
\end{figure}

Solving the DMFT and DCA equations requires the solution of a quantum impurity model. 
Continuous-time quantum impurity solvers
\cite{Rubtsov05,Werner06,Gull08_ctaux}, in particular the
continuous-time auxiliary field (CT-AUX) method used here
\cite{Gull08_ctaux}, combined with advanced numerical techniques
\cite{Gull10_submatrix}, have made it feasible to solve such models
efficiently and numerically exactly on large clusters,
thereby providing a good starting
point for an extrapolation of finite size clusters to the infinite
system
~\cite{Maier05_dwave,Kent05,Kozik10}.
We have performed extensive  DCA calculations on bipartite clusters with
$N_c=18,26,36,48,56$, and $64$. In order to achieve an optimal scaling
behavior we exclusively use the clusters determined in Ref.~\cite{Kent05}
following the criteria proposed by \cite{Betts}. 
As the DCA exhibits a $1/L^2$ finite-size scaling in the
linear cluster size $L=N_c^{1/3}$ \cite{Maier02}, we
extrapolate our cluster results linearly in $N_c^{-2/3}$. Our error bars include extrapolation uncertainties.
Despite a sign problem away from half-filling, temperatures $T/t \ge0.4$, on the order of the N\'eel temperature, are reliably accessible for all  but the largest interaction strength $U=12t$ where  we have been restricted to $T/t\ge0.5$.  

The potential energy, double-occupancy, and
nearest-neighbor spin-spin correlation have been measured directly. The
kinetic energy $E_{\rm kin} = \sum_{n,\vec{k}}\epsilon(\vec{k}) G(\vec{k},i\omega_n)$ has been
calculated by summing $\epsilon(\vec{k})$, the bare-dispersion of the simple
cubic lattice, and the single-particle Green function $G(\vec{k},
i\omega_n)$ over all momenta $\vec{k}$ and Matsubara frequencies $i\omega_n$.
The entropy $S$ has subsequently been calculated by numeric integration
\begin{equation}
S(T) = S(T_{u}) - \frac{E(T_{u})}{T_{u}} + \frac{E(T)}{T}  - \int_{T}^{T_{u}} \!dT' \frac{E(T')}{T'^{2}},
\end{equation}
up to a $T_u/t \approx 10$ ,
where the entropy $S(T_{u})$ is accurately given by a high-temperature
series expansion. 
Tables of the complete results containing finite cluster and extrapolated
values at and away from half filling for the entropy, energy, density, double occupancy, and spin correlations are given in the supplementary material \cite{EPAPS}.

{\it Results at half filling} -- We start our analysis at half
filling and focus on $U/t=8$ where we can compare with results from DMC
simulations~\cite{Staudt00} (and the more accurate ~\cite{ Burovski08}) .
We see in Fig.~\ref{fig:entropy_U8_0} that the entropy calculated
using DCA and DMC coincides within error bars at all temperatures. Agreement
with a 10th order high temperature series expansion \cite{Scarola09} is found down to $T/t \approx 1.6$. At that temperature
also single site DMFT starts to deviate because that method misses short-range antiferromagnetic correlations.
The N{\'e}el temperature was found to be $T_N/t \approx 0.36(2)$ in Ref.~\cite{Kent05}. Our DMC calculations find it more accurately at  $T_N/t = 0.333(7)$.
Using DCA calculations, the critical entropy is $s \approx 0.46(2)$ for $T_N$ according to Ref.~\cite{Kent05}, and $s :=S/N_c \approx 0.41(3)$ with $T_N$ according to the DMC.
In the rest of the paper we will only use the $T_N$ as determined by DMC but with entropies calculated by DCA (since away from half filling only entropies calculated by DCA are available). 


\begin{figure}[t]
  \includegraphics[angle=0, width=0.9\columnwidth]{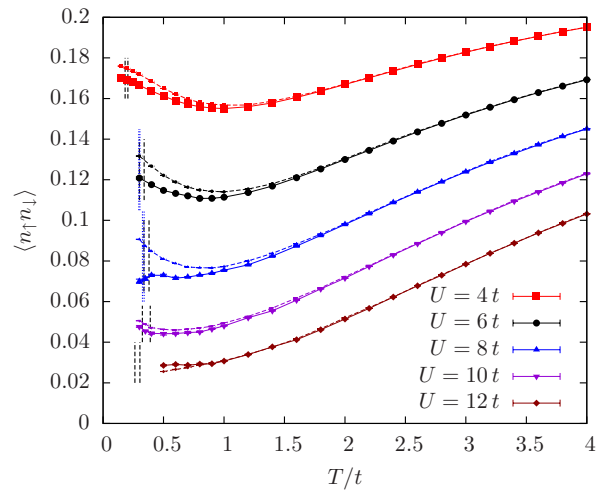}
  \caption{(color online) Double occupancy  of the Hubbard model as a
    function of temperature $T$, at half filling. Extrapolated DCA results are shown as solid lines and  DMFT values as dashed lines. The
    vertical lines are the same as in Fig.~\ref{fig:entropy_U8_0}.  }
  \label{fig:doubleocc_0}
\end{figure}
The \emph{double occupancy}, which has played a crucial role in optical
lattice experiments ~\cite{Joerdens08, Schneider08, Scarola09,
  Strohmaier10}, is shown in Fig.~\ref{fig:doubleocc_0} as a function
of temperature at half filling for different values of $U/t$.  While
for small $U/t$ a remarkable increase is seen on approach to
$T_N$, only a plateau remains at moderate values of
$U/t$. This is in contrast to the DMFT predictions, but similar to
lattice QMC results in two dimensions \cite{Pavia10}. For larger
interactions ($U/t\gtrsim 12$), the double occupancy rises above that of
a single site paramagnet, consistent
with DMFT results for the anti-ferromagnetic phase below
$T_N$\cite{Gorelik}. The negative slope of $D(T)$, discussed in the
context of single site DMFT \cite{Werner05}, persists for a wide range
of parameters.  Sharp features just above $T_N$, as detected in single
site (momentum independent) studies~\cite{Gorelik}, are not observed for the interaction
values and temperature ranges studied here. Hence the proposal that the
double occupancy is a good candidate for thermometry is not substantiated by more
accurate momentum-dependent calculations.

\begin{figure}[t]
  \includegraphics[angle=0, width=0.9\columnwidth]{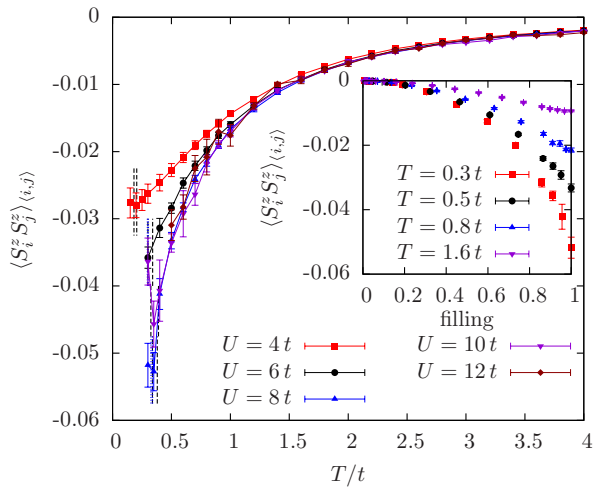}
  \caption{\label{fig:spincorrelation}(color online) Nearest-neighbor
    spin-spin correlation of the Hubbard model as a function of
    temperature $T$, at half filling. The inset shows the
    density dependency for $U/t=8$ and selected
        temperatures. Vertical lines: compare to Fig.~\ref{fig:entropy_U8_0}.}

\end{figure}

The \emph{spin-spin correlation} function plotted in
Fig.~\ref{fig:spincorrelation} as a function of temperature for
various $U$ and as a function of filling for $T/t=0.3, 0.8,$ and
$1.6$ at $U/t=8$, is only accessible in methods that include non-local
correlations, but may be accessible experimentally~\cite{Trotzky2010}. It has a steep slope on approach to the N{\'e}el
temperature, which makes it an ideal quantity for thermometry.
This corresponds to the intuitive picture that charge degrees of
freedom are already essentially frozen out around $T_N$ while the spin degrees
of freedom start to order there.

\begin{figure}[t]
\includegraphics[angle=0, width=0.9\columnwidth]{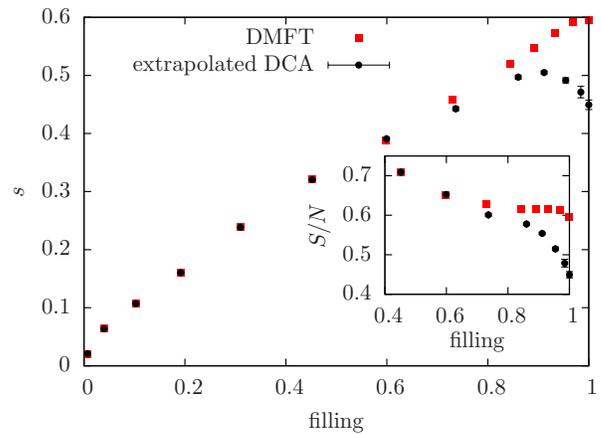}
\caption{(color online) Entropy per lattice site $s$ and the entropy
  per particle $S/N$ (inset) of the Hubbard model
  at the temperature $T=0.35t\approx T_N$, as a function of density $n$,
  for $U/t=8$. DMFT values and extrapolated DCA values
  are shown. The errors are dominated by extrapolation errors.}
\label{fig:entropy_off}
\end{figure}

{\it Results away from half filling} -- Fig.~\ref{fig:entropy_off}
shows the entropy per lattice site for $U/t=8$. The inset demonstrates that entropy per particle number $N$ increases strongly at lower densities. 
While  single site DMFT
remains accurate for densities lower than $n \lesssim 0.6$ due to the
weak momentum dependence of the self-energy in this regime \cite{Gull10_momdep}, the DCA results are important for larger densities.
Similarly, near half
filling DMFT overestimates the double occupancy by $10 \%$, while
deviations are less pronounced at lower densities.  This observation
persists for all interactions and temperatures investigated.  
The flattening of the double occupancy for $8 \le U/t \le 12$
(cf. Fig.~\ref{fig:doubleocc_0}) is also seen away from half filling,
and leads to virtually unchanged profiles over the trap in an optical
lattice system. On the other hand, the spin-spin correlation function
away from half filling (see the inset in
Fig.~\ref{fig:spincorrelation}) changes most rapidly near half filling
when approaching the N{\'e}el temperature since it couples strongly to
the developing (short-range) spin correlations. 

{\it Entropy in the optical lattice system} -- We now turn to the
experimentally relevant case of an optical lattice in a harmonic trap, which is a closed
system where entropy is conserved and temperature changes when
adiabatically changing the parameters of the Hamiltonian. We choose parameters close to current experiments: $V(\vec{r}) = 0.004 (|\vec{r}|/a)^2 t$ with lattice spacing $a$, and we consider the case of half filling in the trap center: $\mu= U/2$.
We treat the harmonic confinement
 in a local density approximation(LDA): for every site we
perform a DCA simulation for a homogeneous system and average the
results over the trap. LDA was found to be a good
approximation for the Bose-Hubbard model for wide traps, except in close proximity to
the critical point~\cite{Wessel04,Campostrini, Pollet2010} of the U(1) phase
transition because of the diverging correlation length and in our setup errors due to the LDA can be neglected compared to our other
systematic errors. 

\begin{figure}[t]
  \includegraphics[angle=0, width=0.9\columnwidth]{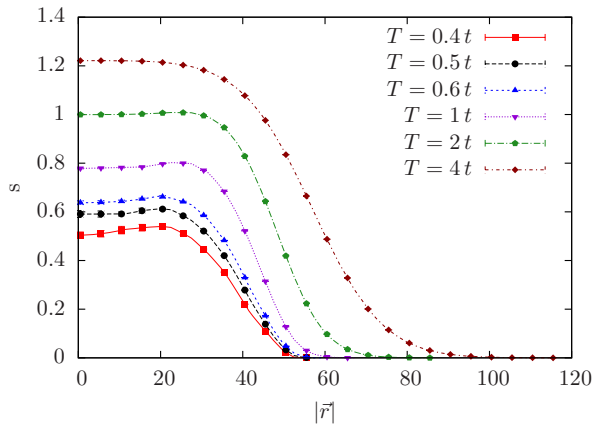}
  \caption{(color online) Entropy profiles (entropy per lattice site)
    plotted over the trap in the LDA approximation for different
    temperatures with an interaction strength $U/t=8$. Error bars are
    shown every 5 lattice spacings, but are generally smaller than the
    symbol size.}
  \label{fig:entropy_profile_U8}
\end{figure}

\begin{figure}[t]
  \includegraphics[width=0.9\columnwidth]{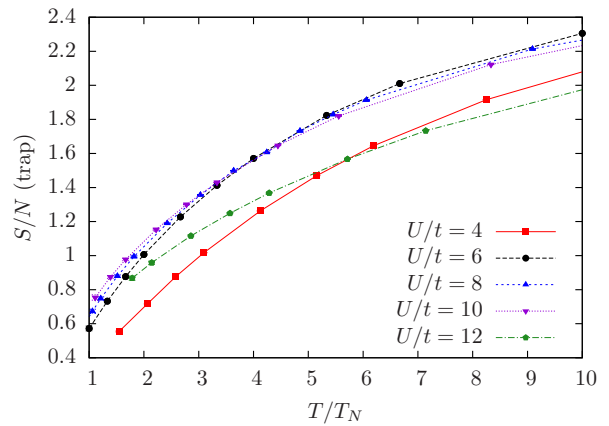}
  \caption{(color online) Entropy per particle averaged over the trap
    as a function of temperature relative to the N{\'eel} temperature
    for different $U$. The N{\'e}el temperature is reached at
    $T/t=0.333(7)$ for $U/t=8$, when the average entropy
    is $S/N = 0.65(6)$. Errors (not shown) in $S(T_N)$ are estimated to be in the
    $10\%$ range, with the largest contribution caused by the
    uncertainty in $T_N$. }
  \label{fig:entropy_temperature}
\end{figure}

Due to the large volume fractions, the wings of the gas may capture more entropy than the center of the trap, even
though the entropy per site is comparable to the one in the center
(see Fig.~\ref{fig:entropy_profile_U8}).  In fact, the entropy
of the whole density range $ 0.1 < n < 0.9$ is large, and this
opens the possibility to observe anti-ferromagnetic order in the trap
center at an average entropy per particle over the trap which is
about $50 \%$ larger than what could be expected  from a homogeneous
study. Optimal parameters are around $U/t=8$ when $T_N/t = 0.333(7)$ according to DMC,
corresponding to $S/N = 0.65(6)$ in the trap, while $S/N=0.41(3)$
would be expected for a homogeneous system. As seen in
Fig.~\ref{fig:entropy_temperature}, all
$U$ in the range $8 < U/t < 12$ lead to similar
conclusions.
We have verified that changing the trap by a factor of 4 does not
alter these conclusions.

{\it Conclusions} -- We have provided the full thermodynamics of the
3d Hubbard model using the DCA formalism for $U/t \le 12$ and
temperatures above the N{\'e}el temperature.   Comparing to single site DMFT results we found that the latter already fail at remarkably high
temperatures ($T/t \approx 1.5$ for $U/t=8$ at the 1\% level) near
half filling.  While the entropy per particle
at the N{\'e}el temperature $T_N/t = 0.333(7)$ (determined with DMC) is
$S/N = 0.41(3)$ for $U/t = 8$ in a homogeneously half filled
system, we
find that the N{\'e}el transition in a trap can already be reached at $S/N = 0.65(6)$
in a realistically sized harmonic trap (taking $T_N$ according to Ref.~\cite{Kent05} leads to $S/N=0.69$).

We have also investigated the double occupancy and the
nearest-neighbor spin-spin correlation function as quantities that are
experimentally measurable and which were suggested to show precursors
of antiferromagnetism. It turns out that the double occupancy
is more or less flat as a function of temperature, while the spin correlations show a strong temperature dependence around the  
N{\'e}el temperature. This suggests that the spin correlations, not the double occupancy, are best suited to observe precursors of antiferromagnetism and measure the temperature. Our numerical data can be used to calibrate such a spin-correlation thermometer.

We acknowledge stimulating discussions with I. Bloch,
T. Esslinger, A. Georges, D. Greif, O. Parcollet, V. Scarola, and L. Tarruell. This work was
supported by the Swiss National Science foundation, the National
Science Foundation grants DMR-0705847 and PHY-0653183,  grant no. ANR-BLAN-6238,  the Aspen Center for Physics, a grant
from the Army Research Office with funding from the DARPA OLE
program, and by the Deutsche Forschungsgemeinschaft
through the collaborative research center SFB~602. We used
the Brutus cluster at ETH Zurich and computational resources
provided by the Norddeutscher Verbund f\"ur Hoch- und
H\"ochstleistungsrechnen (HLRN) and by the Gesellschaft f\"ur
wissenschaftliche Datenverarbeitung G\"ottingen
(GWDG). Simulation codes were based on the ALPS libraries \cite{ALPS}.


\begin{thebibliography}{99}
\bibitem{Hubbard63} J. Hubbard, Proc. Roy. Soc. (London), Ser. A {\bf 276}, 238 (1963).
\bibitem{PWA} P. W. Anderson, Science 235, 1196 (1987).
\bibitem{Imada98} M. Imada, A. Fujimori, and Y. Tokura, Rev. Mod. Phys. 70, 1039 (1998). 
\bibitem{Koehl05} M. K{\"o}hl {\it et al.}, Phys. Rev. Lett. {\bf 94}, 080403 (2005).
\bibitem{Joerdens08} R. J\"ordens {\it et al.}, Nature (London) {\bf 455}, 204 (2008).
\bibitem{Schneider08} U. Schneider {\it et al.}, Science {\bf 322}, 1520 (2008).
\bibitem{DeLeo08} L. De Leo {\it et al.}. Phys. Rev. Lett. {\bf 101}, 210403 (2008).
\bibitem{Scarola09} V.~W. Scarola {\it et al.}, Phys. Rev. Lett. 102, 135302 (2009).
\bibitem{Joerdens10} R. J{\"o}rdens {\it et al.}, Phys. Rev. Lett. {\bf 104}, 180401 (2010).
\bibitem{PotthoffNolting} M. Potthoff and W. Nolting, Phys. Rev. B {\bf 59}, 2549 (1999). 
\bibitem{Gorelik} E. V. Gorelik {\it et al.}, arXiv:1004.4857 (2010).
\bibitem{Kozik10} E. Kozik {\it et al.}, , Europhys. Lett. {\bf 90}, 10004 (2010).
\bibitem{Trotzky09} S. Trotzky {\it et al.}, arXiv:0905.4882v1, to appear in Nature Physics.
\bibitem{Maier05} T. Maier {\it et al.}, Rev. Mod. Phys. {\bf 77}, 1027 (2005).
\bibitem{Hettler98} M. H. Hettler {\it et al.}, Phys. Rev. B {\bf 58}, R7475 (1998).
\bibitem{Georges96} A. Georges {\it et al.}, Rev. Mod. Phys. {\bf 68}, 13 (1996).
\bibitem{Rubtsov05}A. N. Rubtsov, V. V. Savkin, and A. I. Lichtenstein, Phys. Rev. B {\bf 72}, 035122 (2005).
\bibitem{Gull08_ctaux} E. Gull {\it et al.}, EPL {\bf 82}, 57003 (2008).
\bibitem{Werner06} P. Werner {\it et al.}, Phys. Rev. Lett. {\bf 97}, 076405 (2006); P.~Werner and A.~J. Millis, Phys. Rev. B {\bf 74}, 155107(2006).
\bibitem{Gull10_submatrix} E. Gull {\it et al.}, in preparation.
\bibitem{Kent05} P. Kent {\it et al.}, Phys. Rev. B {\bf 72}, 060411(R) (2005).
\bibitem{Maier05_dwave} T. Maier {\it et al.}, Phys. Rev. Lett. 95, 237001 (2005);
\bibitem{Betts} D. D. Betts and G. E. Stewart, Can. J. Phys. {\bf 75}, 47 (1997).
\bibitem{Maier02} T. Maier and M. Jarrell, Phys. Rev. B {\bf 65}, 041104 (2002).
\bibitem{EPAPS} see accompanying EPAPS document no. and the {\tt aux} folder of the arXiv sources.
\bibitem{Staudt00} R. Staudt, M. Dzierzawa, and A. Muramatsu, Eur. Phys. J. B {\bf 17}, 411 (2000).
\bibitem{Burovski08} E. Burovski {\it et al.}, Phys. Rev. Lett. {\bf 101}, 090402 (2008).
\bibitem{Werner05} F. Werner {\it et al.}, Phys. Rev. Lett. {\bf 95}, 056401 (2005).
\bibitem{Strohmaier10} N. Strohmaier {\it et al.}, Phys. Rev. Lett. {\bf 104}, 080401 (2010).
\bibitem{Pavia10} T. Pavia {\it et al.}, Phys. Rev. Lett {\bf 104}, 066406 (2010).
\bibitem{Trotzky2010} S. Trotzky {\it et al.}, arXiv:1009.2415 (2010).
\bibitem{Gull10_momdep} E. Gull {\it et al.}, arXiv:1007.2592 (2010).
\bibitem{Wessel04} S. Wessel {\it et al.}, Phys. Rev. A  {\bf 70}, 053615 (2004).
\bibitem{Campostrini} M. Campostrini and E. Vicari, Phys. Rev. Lett. {\bf 102}, 240601 (2009); Phys. Rev. A, {\bf 81}, 023606 (2010).
\bibitem{Pollet2010} L. Pollet, N. V. Prokof'ev, and B. V. Svistunov, Phys. Rev. Lett. {\bf 104}, 245705 (2010).
\bibitem{ALPS} A. F. Albuquerque,  {\it et al.}, Journal of Magnetism and Magnetic Materials {\bf 310}, 1187 (2007).
\end{thebibliography}
\end{document}